\documentclass[twocolumn,pra,showpacs,floatfix]{revtex4}
\usepackage{amsmath}
\usepackage{graphicx}
\newcommand{\ket}[1]{\ensuremath{|#1\rangle}}

\newcommand{\nuc}[2]{\mbox{${}^{#1}\rm #2$}}

\newcommand{\NOT}{\textsc{not}}

\newcommand{\CNOT}{controlled-\NOT}

\begin{document}
\title{Robust Logic Gates and Realistic Quantum Computation}
\author{Li Xiao}
\author{Jonathan A. Jones}\email{jonathan.jones@qubit.org}
\affiliation{Centre for Quantum Computation, Clarendon Laboratory,
University of Oxford, Parks Road, OX1 3PU, United Kingdom}
\date{\today}
\begin{abstract}
The composite rotation approach has been used to develop a range of
robust quantum logic gates, including single qubit gates and two
qubit gates, which are resistant to systematic errors in their
implementation. Single qubit gates based on the BB1 family of
composite rotations have been experimentally demonstrated in a
variety of systems, but little study has been made of their
application in extended computations, and there has been no
experimental study of the corresponding robust two qubit gates to
date. Here we describe an application of robust gates to Nuclear
Magnetic Resonance (NMR) studies of approximate quantum counting. We
find that the BB1 family of robust gates is indeed useful, but that
the related NB1, PB1, B4 and P4 families of tailored logic gates are
less useful than initially expected.
\end{abstract}
\maketitle

Quantum information processing \cite{bennett00} has made substantial
progress in recent years, but formidable problems remain in the
implementation of a general purpose quantum computer.  The
development of quantum error correction \cite{shor95, steane96,
steane99} was a key step, as it allows quantum computers to function
in the presence of random errors.  More recently the method of
composite rotations (also called composite pulses), developed for
Nuclear Magnetic Resonance (NMR) experiments \cite{ernst87,
freeman97b, levitt86, wimperis94}, has been used to design quantum
logic gates which are robust against \textit{systematic} errors in
the control fields used to implement them \cite{cummins00,
cummins02}.

Composite pulses are closely related to the more general field of
optimal quantum control, which has been widely studied and finds
applications in many fields.  In the context of conventional NMR
experiments there has been much study of the design of time-optimal
pulse sequences \cite{khaneja05b}, which enable unitary
transformations to be performed while minimizing losses from
decoherence.  For applications in NMR quantum computing there has
been particular interest in the use of strongly modulated composite
pulses \cite{fortunato02, boulant03, weinstein04}, which are
designed to perform particular selective operations in the presence
of a complex multi-spin Hamiltonian, and which can also be designed
to be robust against systematic errors in control fields. Although
often discussed separately there are clear similarities between
these approaches \cite{khaneja05a}.  Similar methods have also been
developed in ion trap implementations of quantum computing
\cite{gulde03}.

Henceforth we only consider a particularly simple group of composite
gates, based on the composite pulses originally developed for
conventional NMR experiments. These pulses differ from the more
general approaches mentioned above in two significant ways. Firstly
they provide a simple analytic set of solutions, which allow simple
rotations to be converted directly into composite rotations, rather
than a recipe for finding a numeric solution for a particular
problem. Secondly, the implementation of these composite gates is
particularly simple.  For example, in NMR systems they only require
control of the phase and duration of a small number of RF pulses at
a single fixed frequency and power, while strongly modulated
composite pulses also require the frequency and power to be varied.
For this reason, these pulses are easy to implement experimentally.

Single qubit gates developed using this approach have been
demonstrated experimentally in a range of systems, including NMR
\cite{cummins00, cummins02}, Electron Spin Resonance (ESR)
\cite{morton05a, morton05b}, and the Quantronium SQUID
\cite{collin04}, but there has not yet been much study of their use
in an extended quantum algorithm. A related family of robust two
qubit gates has also been described \cite{jones03a, jones03b}, but
these particular gates have not so far been studied experimentally.
Here we described the application of single and two qubit logic
gates based on the BB1, NB1 and PB1 families of composite pulses to
an implementation of approximate quantum counting using NMR studies
of a heteronuclear spin system.

\section{The BB1, NB1 and PB1 families}

Composite pulses are designed to perform quantum operations in the
presence of systematic errors.  Note that it is not necessary to
know the size of the error involved, but it is necessary to know its
general form.  They have been extensively developed in conventional
NMR studies \cite{levitt86}, principally to correct \textit{pulse
length errors}, that is errors in the strength of the field used to
induce a rotation, and \textit{off-resonance errors}, that is
imperfections arising when an oscillating field is not quite in
resonance with the transition it is supposed to drive.  Most of
these composite pulses are not suitable for use in quantum
information processing, but a small group of them, sometimes called
fully compensating pulses, are well suited.  These pulses work on
any initial state, and can in principle be used to replace naive
gates without any further modifications.

Perhaps the most useful family of fully compensating pulses
developed to date is the BB1 family of pulses which are robust
against pulse length errors.  The ideal unitary transformation for a
$\theta_\phi$ pulse is
\begin{equation}
U(\theta,\phi)=\exp[-i\theta(I_x\cos\phi+I_y\sin\phi)]
\end{equation}
but in a real physical system the actual operation will not have
this perfect form.  In the presence of pulse length errors the real
operation has the form
\begin{equation}
V(\theta,\phi)=\exp[-i(1+f)\theta(I_x\cos\phi+I_y\sin\phi)]
\end{equation}
where $f$ is the fractional error in the pulse length or field
power. Defining the propagator fidelity as
\begin{equation}
F=|Tr(VU^\dag)|/Tr(UU^\dag)
\end{equation}
gives the result
\begin{equation}
F=\cos(\frac{f\theta}{2})\approx1-\frac{f^2\theta^2}{8}
\end{equation}
for a simple $\theta_\phi$ pulse.  The BB1 family of robust pulses
is obtained by replacing the simple pulse $\theta_\phi$ by the
composite pulse sequence
\begin{equation}
(\theta/2)_\phi (\pi)_{\phi+\psi} (2\pi)_{\phi+3\psi}
(\pi)_{\phi+\psi} (\theta/2)_\phi \label{eq:BB1}
\end{equation}
where $\psi=\arccos(-\theta/4\pi)$. This composite pulse has a
fidelity of
\begin{equation}
F\approx1-f^6\frac{(32\pi^4\theta^2+14\pi^2\theta^4-\theta^6)}{9216}
\end{equation}
showing that for small errors the BB1 pulse is far less error prone
than a simple pulse.

The NB1 family of pulses is obtained in a similar way; in this case
the simple pulse is replaced by
\begin{equation}
(\theta/2)_\phi (\pi)_{\phi+\psi} (2\pi)_{\phi-\psi}
(\pi)_{\phi+\psi} (\theta/2)_\phi
\end{equation}
which has the same structure as BB1, differing only in the phases.
This sequence is \textit{more} error-prone than a simple
$\theta_\phi$ pulse, and so cannot be used to suppress errors. It
does, however, have the interesting property that when $|f|\approx1$
the resulting composite pulse is a good approximation to the
Identity operation.  Thus is the case of very low pulse powers
($f\approx-1$) an NB1 pulse ``does nothing'': in effect, evolution
under weak fields is suppressed.

The PB1 family of pulses offers a compromise between these two
extremes.  The simple pulse is replaced by
\begin{equation}
(\theta/2)_\phi (2\pi)_{\phi+\psi'} (4\pi)_{\phi-\psi'}
(2\pi)_{\phi+\psi'} (\theta/2)_\phi
\end{equation}
where $\psi'=\arccos(-\theta/8\pi)$.  These pulses are both robust
to small errors \textit{and} able to suppress weak fields,
suggesting that PB1 is the best general purpose family of pulses,
although the BB1 and NB1 families perform their respective tasks
more effectively.

Since these families were described, a range of similar pulses have
been discovered, most notably the arbitrary precision pulses of
Brown \textit{et al.} \cite{brown04, brown05}.  In their notation
BB1 is called B2 and PB1 is called P2; these are the first members
of a series of families of pulses with ever greater error tolerance.
These sequences swiftly become extremely long, and here we consider
only the next two members, B4 and P4.  The B4 composite pulse takes
the form
\begin{multline}
(\theta/2)_\phi
\left[(\pi)_{\phi+\psi} (2\pi)_{\phi+3\psi} (\pi)_{\phi+\psi}\right]^4\\
(-2\pi)_{\phi+\psi}(-4\pi)_{\phi-\psi}(-2\pi)_{\phi+\psi}\\
\left[(\pi)_{\phi+\psi} (2\pi)_{\phi+3\psi} (\pi)_{\phi+\psi}\right]^4
(\theta/2)_\phi
\end{multline}
where $\psi=\arccos(-\theta/24\pi)$ and the superscript 4 indicates
that the section enclosed by square brackets is repeated four times.
The central three pulses have negative rotation angles; these can be
partially canceled with surrounding pulses, and the remaining
rotations can be implemented as positive rotations around axes with
phase angles offset by $\pi$. The P4 composite pulse is similar,
taking the form
\begin{multline}
(\theta/2)_\phi
\left[(2\pi)_{\phi+\psi'} (4\pi)_{\phi-\psi'}(2\pi)_{\phi+\psi'}\right]^4\\
(-4\pi)_{\phi+\psi'}(-8\pi)_{\phi-\psi'}(-4\pi)_{\phi+\psi'}\\
\left[(2\pi)_{\phi+\psi'} (4\pi)_{\phi-\psi'}(2\pi)_{\phi+\psi'}\right]^4
(\theta/2)_\phi
\end{multline}
where $\psi'=\arccos(-\theta/48\pi)$. Note that the forms for these
pulses originally published by Brown \textit{et al.} \cite{brown04}
are slightly incorrect \cite{brown05}.

Under ideal conditions B4 and P4 perform slightly better than their
simpler counterparts, but the gain is slight and more detailed
simulations \cite{naylor} indicate that these pulses are highly
sensitive to the presence of off-resonance and phase errors.  This
is confirmed by our experience, described below, that the
performance of B4 and P4 is in practice quite poor. Thus the
original three families offer perhaps the best combination of
simplicity and effectiveness, and we concentrate on them in most of
what follows. It might seem that, as previously suggested, PB1 is
the best general purpose family of pulses, but this turns out not to
be the case.

\section{The quantum counting experiment}
Most studies of these particular robust quantum logic gates to date
have been either theoretical or have involved only simple
demonstrations of single logic gates. Although these are of some
interest, it is also important to investigate more complex
situations, such as quantum algorithms containing large numbers of
logic gates.  Quantum counting provides an ideal testing ground as
it permits large numbers of logic gates to be implemented in a
simple physical system.

\begin{figure}
\begin{center}
\includegraphics[scale=1]{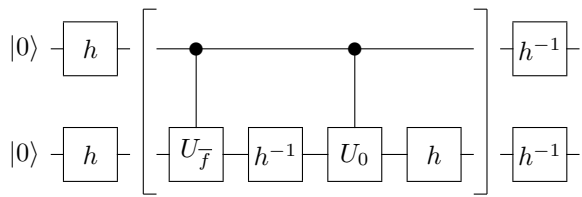}
\end{center}
\caption{A quantum circuit for implementing quantum counting on a
two qubit NMR quantum computer; the central sequence of gates,
surrounded by brackets, is applied $r$ times.  A similar circuit can
be constructed for a larger search space by replacing the (lower)
target bit by a register and replacing gates applied to the target
by multi-bit versions.  Gates marked $h$ implement the NMR
pseudo-Hadamard operation, while those marked $h^{-1}$ implement the
inverse operation.} \label{fig:circuit}
\end{figure}

Quantum counting is closely related to Grover's quantum search
\cite{grover97, grover98, jones98b}. Consider a function $f(x)$
which maps $n$-bit strings to a single output bit, such that
$f(x)=0$ or $1$. In general there are $N=2^n$ possible input values,
with $k$ values for which $f(x)=1$. Grover's quantum search allows
one of these $k$ items to be located, while quantum counting
\cite{boyer98, jones99} allows the value of $k$ to be estimated. The
counting algorithm estimates an eigenvalue of the Grover iterate $G
= H U_0 H^{-1} U_{\overline{f}}$, which forms the basis of the
searching algorithm, where $H$ is the $n$-bit Hadamard transform,
$U_0$ maps $|000\ldots0\rangle$ to $-|000\ldots0\rangle$ and leaves
the remaining basis states alone, and $U_{\overline{f}}$ maps
\ket{x} to $(-1)^{f(x)+1}\ket{x}$.  For further details see
\cite{jones99}.

A quantum circuit which implements this algorithm on a two qubit NMR
quantum computer is shown in Fig.~\ref{fig:circuit}.  As usual,
pairs of Hadamard gates have been replaced by NMR pseudo-Hadamard
gates and their inverse \cite{jones99, jones98}.  This circuit can
be used to count the number of solutions to $f(x)=1$ over a one bit
search space, but similar circuits exist for larger search spaces.

\section{Single qubit gates}
To investigate the effects of systematic errors on single qubit
gates we implemented this circuit in an NMR experiment using a
solution of \nuc{13}{C} labeled sodium formate in
$\text{D}_2\text{O}$ \cite{xiao05}.  The formate anion
$\text{HCO}_2^-$ provides an isolated two spin system, comprising a
\nuc{1}{H} and a \nuc{13}{C} nucleus, with a large coupling between
the nuclei.  This coupling arises from the scalar coupling
interaction, and in a heteronuclear spin system takes the
weak-coupling form, usually written as $\pi J\,2I_zS_z$ in NMR
product operator notation \cite{ernst87}.  (This coupling is
sometimes said to have the Ising form, although this name more
properly refers to extended networks of such couplings.) In a
heteronuclear spin system the RF transmitters can be place in
perfect resonance with each qubit, and so off-resonance effects can
be essentially ignored. The major remaining source of systematic
error is pulse length errors arising from inhomogeneity of the RF
fields.

\begin{figure}
\begin{center}
\includegraphics[scale=0.4]{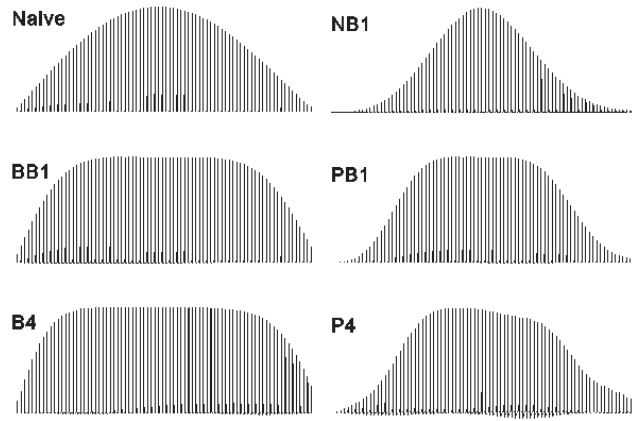}
\end{center}
\caption{The effects of pulse length errors on a single qubit
pseudo-Hadamard gate (a $90^\circ_y$ rotation).  The figures show
the effect of applying this gate to a spin in the thermal
equilibrium state, which is equivalent to a pseudo-pure state
\ket{0}, followed by observation of the NMR spectrum, effectively
measuring the off-diagonal components of the single-spin density
matrix.  For further details see the main text.}
\label{fig:hadamard}
\end{figure}
\begin{figure*}
\begin{center}
\includegraphics[scale=0.7]{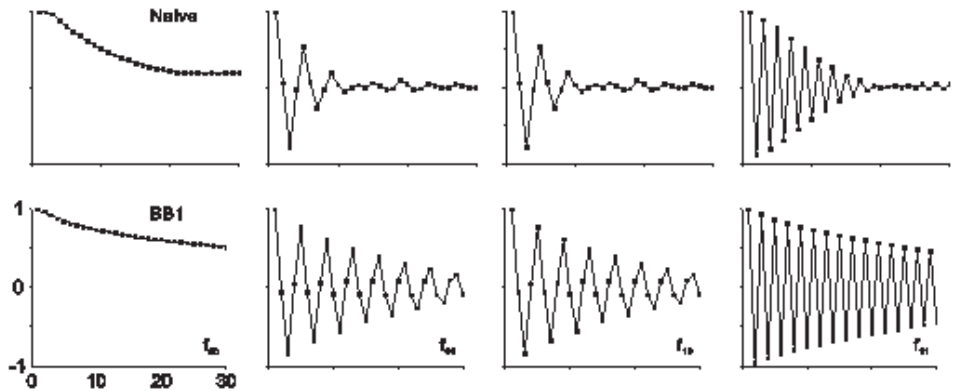}
\end{center}
\caption{An implementation of approximate quantum counting with
naive and BB1 single qubit gates.  The use of BB1 single qubit gates
greatly reduces the apparent decoherence rate, indicating that much
of the signal loss actually arises from pulse length errors, and
that BB1 is effective in correcting for this in a complex sequence.}
\label{fig:countres}
\end{figure*}
We begin by demonstrating the effect of using robust quantum logic
gates to implement a pseudo-Hadamard gate (a $90^\circ_y$ rotation)
on the \nuc{1}{H} qubit which begins in an equilibrium state.  In
addition to the pulse length errors which arise naturally from
inhomogeneity, additional errors can be introduced by deliberately
mis-setting the pulse length.  The effect of doing this is shown in
Fig.~\ref{fig:hadamard}.  Each sub-figure contains a set of spectra
corresponding to pulse length errors in the range $\pm100\%$, while
the six sub-figures correspond to naive pulses and the five families
of composite pulses described previously.

As expected the naive pulse gives a signal whose intensity is a
cosine function of the fractional error. The BB1 family is much more
robust to pulse length errors, while the NB1 family is more
sensitive to errors than the naive pulse, and gives much smaller
excitations than the naive pulse for errors close to $\pm100\%$. The
PB1 family shows the expected compromise behaviour, with a broad
central maximum for small errors, and broad minima around errors of
$\pm100\%$. Note that the performance is better for errors around
$-100\%$, which is the experimentally important case of very weak
fields, than for the theoretically equivalent but experimentally
unimportant case of errors close to $+100\%$; this reflects the
intrinsic inhomogeneity of the RF field.  The performance of the B4
composite pulse is broadly similar to that of the much simpler BB1
sequence, while the performance of P4 is clearly rather poor.

This shows that the robust logic gates can have the desired effects
when used to implement a single logic gate, with BB1 and PB1
offering the best performance, but it is also important to
investigate how they work in more complex situations.  This was done
by implementing the quantum counting circuit with the results shown
in Fig.~\ref{fig:countres}. Only naive and BB1 single qubit gates
were used. The improvement obtained by using BB1 single qubit gates
is clear, with much of the signal loss that would naively be
ascribed to decoherence clearly arising from pulse length errors.
BB1 single qubit gates are, therefore, used throughout the remainder
of this paper.

\section{Two qubit gates}
\begin{figure}
\begin{center}
\includegraphics{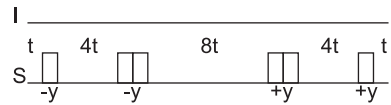}
\end{center}
\caption{Pulse sequence for a BB1 robust coupling gate to implement
a \CNOT\ gate in a two spin ($IS$) system. Boxes correspond to
single qubit rotations with rotation angles of
$\psi=\arccos(-1/8)\approx97.2^\circ$ applied along the $\pm y$ axes
as indicated; time periods correspond to free evolution under the
scalar coupling, $\pi J\,2I_zS_z$, for multiples of the time
$t=1/4J$. The naive coupling gate corresponds to free evolution for
a time $2t$.} \label{fig:pulses}
\end{figure}

A very similar approach can also be used to tackle systematic errors
in coupling gates, which provide the basic two qubit gate for NMR
quantum computation \cite{jones98}. Evolution under a scalar
coupling can be thought of as a rotation about the $2I_zS_z$ axis,
and errors in the coupling constant $J$ correspond to errors in a
rotation angle about this axis.  Such errors can be overcome
\cite{jones03a, jones03b} by rotating about a sequence of axes
tilted from $2I_zS_z$ towards another axis, such as $2I_zS_x$.
Defining
\begin{equation}
\theta_\phi\equiv\exp[-i\times\theta\times(2I_zS_z\cos\phi+2I_zS_x\sin\phi)]
\end{equation}
allows the naive sequence $\theta_0$ to be replaced by
Eq.~\ref{eq:BB1} as before. The tilted evolutions can be achieved by
sandwiching free evolution under the coupling Hamiltonian between
$\phi_{\mp y}$ pulses applied to spin $S$.  For the case that
$\theta=\pi/2$ (which forms the basis of the \CNOT\ gate) the final
sequence takes the form shown in Fig.~\ref{fig:pulses}.

The results of using these robust two qubit gates could be
investigated in much the same way as the single qubit gates shown
above, but it is more interesting to consider another phenomenon,
that is the ability of the NB1 and PB1 gates to suppress evolution
under small scalar couplings.  This was studied by implementing
quantum counting on the more complex spin system provided by alanine
which is  \nuc{13}{C} labeled at position 2.  Like the formate
anion, this contains a CH spin system, comprising the labeled
\nuc{13}{C} and the directly bonded \nuc{1}{H} nucleus, but unlike
formate this spin system is not completely isolated, as both nuclei
have small couplings to the \nuc{1}{H} nuclei in the methyl group.
This additional coupling interaction could be removed by deuterating
the methyl group, or by applying decoupling fields, but here we seek
to suppress the coupling using robust logic gates.

\begin{figure*}
\begin{center}
\includegraphics[scale=0.7]{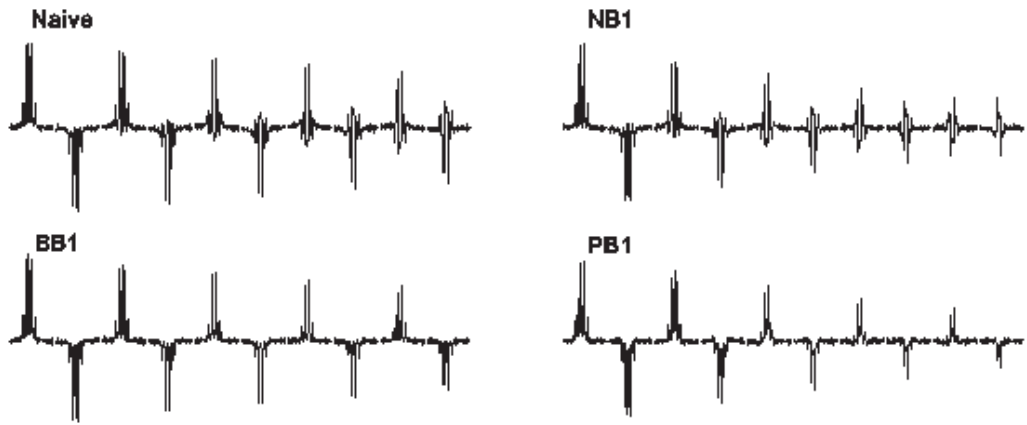}
\end{center}
\caption{Implementation of an coupling gate in the presence of
extraneous couplings.  For details see the main text.  The ten
\nuc{13}{C} spectra in each subfigure correspond to increasing
periods of evolution under the coupling, and the desired pattern is
a simple alternation in signal intensity between $+1$ and $-1$;
deviations in the naive implementation arise from evolution under
the extraneous couplings to the methyl protons. It might be expected
that evolution under this small coupling would be effectively
suppressed by the NB1 sequence, but this performs very poorly, and
the best results are seen from the BB1 sequence.} \label{fig:Ising}
\end{figure*}

We begin by studying evolution under the CH coupling, and
concentrate on evolution times of $n/2J$, where $n$ is a positive
integer.  This was done by applying a pseudo-Hadamard gate to the
\nuc{13}{C} nucleus and then allowing the spin system to evolve
under a coupling gate (either naive or composite) for an effective
time $n/2J$.  The results are shown in Fig.~\ref{fig:Ising}.  The
naive gate appears to perform well for small evolution times, as the
evolution under the small $J$ coupling can be largely ignored: this
coupling is clearly visible in the spectra where each component of
the doublet is split into a $1:3:3:1$ multiplet, but all four
components appear to have very similar phases. For longer evolution
times, however, the detrimental effect of this additional coupling
becomes clear, as the components appear with quite different phases,
indicating significant evolution under the coupling.

In an attempt to overcome this we implemented NB1 coupling gates,
which should suppress evolution under this small coupling.  It is
immediately obvious from Fig.~\ref{fig:Ising} that this approach
does not work, as the multiplets are far more distorted than those
seen using the naive sequence.  In retrospect this is unsurprising:
NB1 is designed to suppress a small coupling on the assumption that
this is the only coupling present. If the small coupling occurs in
addition to another much larger coupling then there is no reason to
believe that NB1 will suppress it.

Examination of the results from BB1 and PB1 coupling gates makes
this point even more clearly.  BB1 gives much cleaner spectra than
the naive coupling gate, with the results of PB1 being similar but
significantly poorer. Once again the explanation is clear in
retrospect: although we know that the splittings visible in the
spectrum arise from the combination of a large coupling and a small
coupling, the pulse sequence is blind to this origin.  The same
pattern could in principle arise from a mixture of different
molecules with a \textit{range} of coupling constants. The BB1
robust gate is designed to give very similar behaviour over this
range of coupling sizes, and so all the different components of the
multiplet appear inphase.  The action of PB1 is similar: the quality
of the spectra seen arises not from the ability of this sequence to
suppress small couplings, but from its ability to tolerate a range
of couplings. Over the range of couplings seen in this system the
behaviour of PB1 and BB1 should be very similar, and the relatively
poor results observed from PB1 are probably a consequence of the
fact that the PB1 sequence take almost twice as long as BB1, leading
to increased problems from decoherence.

\section{Simplifying spin systems}
The results above show that our original idea that NB1 based
composite two qubit gates could be used to suppress small couplings,
in effect simplifying complex spin systems, is incorrect.  However,
BB1 based two qubit gates do result in the desired suppression, with
PB1 sequences giving similar but slightly poorer results, suggesting
that it might be possible to use these to simplify spin systems
instead.

Following this idea we attempted to implement quantum counting in
our labeled alanine spin system, using BB1 to suppress the small
couplings between the \nuc{13}{C} qubit and the methyl protons while
using the large coupling between the \nuc{13}{C} and the directly
bonded \nuc{1}{H} qubit to implement logic gates.  These attempts
were not successful (data not shown), and once again it is clear in
retrospect why this idea will not work. Although the small couplings
between the \nuc{13}{C} and the methyl protons in the labeled
alanine spin system can indeed be suppressed by BB1, there are also
small couplings between the \nuc{1}{H} qubit and the methyl protons
which will not be suppressed.  These homonuclear couplings could, of
course, be removed by frequency selective pulses, or by decoupling,
but either approach would act to directly simplify the spin system,
rendering the BB1 approach unnecessary.

\section{Conclusions}
Our results confirm that simple composite pulses can indeed be used
to suppress systematic errors on single qubit gates used in
implementations of complex quantum algorithms.  The BB1 approach is
likely to prove the most useful, while PB1 may find applications in
some special areas.  The BB1 sequence can also be used to implement
robust two qubit gates, although in this case the extra time
required for the longer pulse sequence may cause difficulties.  The
more complex B4 and P4 sequences, although theoretically superior,
do not perform well in practice.

Two qubit gates based on NB1 and PB1 could in principle be used to
suppress the effects of small couplings, but this is not effective
when the small couplings occur in addition to larger couplings which
are used to implement gates.  In this case the small couplings can
be treated as small errors in the large coupling, and BB1 provides
the best suppression of their effects.  It might appear that this
approach could be used to neglect small couplings in complex spin
systems, effectively simplifying the spin system through composite
pulses, but this approach will rarely if ever be effective, as
corresponding homonuclear couplings cannot be suppressed by these
simple composite pulse methods.  More complex approaches, such as
strongly modulated composite pulses, could be more effective, but
such pulses need to be designed on a case by case basis.

\begin{acknowledgments}
We thank the UK EPSRC and BBSRC for financial support and P.~Baker,
K.~R. Brown and W.~Naylor for helpful conversations.
\end{acknowledgments}

\end{document}